\documentclass{article}
\usepackage{graphicx}
\usepackage{amsmath}

\begin{document}

\bigskip \bigskip
\begin{titlepage}
\title{\bf  Asymptotics of Quantum Random Walk Driven by Optical
Cavity\footnote{Journal of Optics B: Quantum Semiclass. Opt. 7
(2005) S152-S157} }
\date{}
\author{
{\bf Demosthenes Ellinas}$^{*}$ and  {\bf Ioannis Smyrnakis} $^{*
}$
\thanks{
ellinas@science.tuc.gr; $^{** }$ smyrnaki@tem.uoc.gr}
\\
Department of Sciences, Division of Mathematics \\
Technical University of Crete GR-731 00 Chania Crete Greece}
\maketitle
\begin{center}
{\bf Abstract}\end{center}
    We investigate a novel quantum random walk (QRW) model, possibly useful in
quantum algorithm implementation, that achieves a quadratically faster diffusion
rate compared to its classical counterpart.

 We evaluate its asymptotic behavior expressed in the form of a limit probability
distribution of a double horn shape. Questions of robustness and control of that limit
distribution are addressed by introducing a quantum optical cavity in which a
resonant Jaynes-Cummings type of interaction between the quantum walk coin
system realized in the form of a two-level atom and a laser field is taking place.
Driving the optical cavity by means of the coin-field interaction time and the initial
quantum coin state, we determine two types of modification of the asymptotic
behavior of the QRW. In the first one the limit distribution is robustly reproduced up
to a scaling, while in the second one the quantum features of the walk, exemplified
by enhanced diffusion rate, are washed out and Gaussian asymptotics
prevail.

Verification of these findings in an experimental setup that involves two quantum
optical cavities that implement the driven QRW and its quantum to classical transition
is discussed.

\end{titlepage}\bigskip

\section{Introduction}

In a quantum random walk (QRW)\cite{kempeReview}, a topic of intense
research activity in the field of Quantum Computing and \ Information during
the last years, the statistical correlations between the random coin and
walker of a classical random walk (CRW), are replaced by quantum
correlations. This is achieved by appropriate quantization of coin-walker
systems and by their dynamic interaction that built up entanglement\cite%
{nielsenchuang} between them in the course of time evolution of the walk.

Some of the main interests in QRW studies have been the effect of
entanglement on various asymptotics, on spreading properties, and on hitting
and mixing times. From the earlier formulations of QRWs \cite{aharonov}\cite%
{mayer}, to recent works on general graphs \cite{ambainis}, on the line \cite%
{daharonov} \textit{etc.}, it has been shown that some surprising features
use to distinguish the quantum from classical walks, these included features
such as non-Gaussian asymptotics\cite{konno}\cite{grimmett}, quadratic speed
up in spreading rate for walks on line \cite{nayak}, exponentially faster
hitting time in hypercubes\cite{moore}\cite{kempe}, exponentially faster
penetration time of decision trees \cite{childs1}\cite{childs2}, breaking of
majorization ordering and subsequent increase of entropy on average of
position probabilities during time evolution of the walk\cite{bet} \textit{\
etc. }On the experimental side of the investigations on QRWs a number of
proposals have been put forward for their implementation e.g in ion traps
\cite{travaglione}, optical lattices \cite{dur}, or in cavity QED \cite%
{sanders}.

In the present work we study a novel model of a QRW the so called $V^{2}$
model, that was introduced in \cite{bet}(see also\cite{ell}). This is a
discrete time homogeneous walk with constant unitary evolution operator $%
V^{2},$ which is the square of some $V$ operators (see below).
Contrary to all other models of QRW, e.g
\cite{ambainis}\cite{travaglione}, that evolve by increasing powers
of the unitary $V,$ i.e $V,V^{2},V^{3}...$ , the present homogeneous
model is known to have probability distributions at each step that
do obey the majorization ordering and have a constant increase in
their degree of mixing (entropy)\cite{bet}. Also this model is
exactly solvable both for its finite and long time asymptotic
probability distributions, and exhibits \ a quadratically faster
spreading rate of its position in comparison to CRW \cite{es}.

The aim of this study is to investigate the effect of changes of the
quantum coin, identified physically with an atom of two levels
modeling the head-tail states of a random coin, upon the asymptotic
probability distribution function (pdf) of the walk. This limit pdf
is obtained and has a distinct double horn shape. The changes on the
coin are induced by letting it to cross an optical cavity where it
interacts with a single mode \ EM field. The type of interaction
implemented in the cavity is the resonant Jaynes-Cummings model
(JCM)\cite{jc}\cite{ss}\cite{colrev}\cite{engl}\cite{reim}. In fact
in order to broad the type of changes made on the quantum coin, we
will solve for the effects of three other solvable variation of the
resonant JCM, namely the
intensity dependent JCM\cite{bs}, the two photon JCM\cite{sb}, and the $m-$%
photon JCM\cite{singh}. Since a cavity QED realization of QRW has already
been proposed\cite{sanders}, we will say that our cavity driving and
controlling the walk that precedes that one is the \textit{first cavity,}
while the cavity implementing the walk is the \textit{second cavity}.

The outline of the paper is as follows: in Chapter 2, the QRW, its
moments and its limit pdf are obtained, in Chapter 3, the effect of
the first cavity in the form of a completely positive trace
preserving map (CPTP), acting on the coin density matrix is
obtained, in Chapter 4, the conditions for which the limit pdf
survives or gets destroyed due to effects of the driving cavity are
obtained and analyzed, and finally in the last Chapter a number of
conclusions are summarized and some of the future extensions of the
work are mentioned.

\section{Quantum Random Walk and its Asymptotics}

Lets us consider a quantum walker system with Hilbert space $%
H_{W}^{N}=span\{\left\vert n\right\rangle |\;n\in Z\}$ and a quantum coin
system with space $H_{C}=span\{\left\vert +\right\rangle ,\left\vert
-\right\rangle \}$. \ Suppose that the initial state of the walker is given
e.g by the density matrix $\rho _{W}=\left\vert 0\right\rangle \left\langle
0\right\vert $ while the initial coin state is a general density matrix $%
\rho _{C}$ defined in $H_{C}.$ The evolution operator is taken to be
\begin{equation}
V=P_{+}U_{0}\otimes E_{+}+P_{-}U_{0}\otimes E_{-},
\end{equation}%
a conditional step operator for the walker state vectors where $U_{0}=e^{i%
\frac{\pi }{4}\sigma _{2}}$ is a $\frac{\pi }{4}-\ $rotation matrix, $E_{\pm
}\left\vert n\right\rangle =\left\vert n\pm 1\right\rangle $ are the
right/left step unitary operators in the walker system, and $P_{+},$ $P_{-}$
\ the projections in the coin space along the $\left\vert +\right\rangle
,\left\vert -\right\rangle $ directions. Also important are two other
operators, the \textit{walker position operator }$L,$ which satisfies the
eigenvalue equation $L|n\rangle =n|n\rangle ,$ and the \textit{phase operator%
} $\Phi $ that acts on the Fourier transformed states $H_{W}^{\Phi
}=span\{\left\vert \phi \right\rangle =\frac{1}{2\pi }\sum_{n\in
Z}e^{-in\phi }\left\vert n\right\rangle $ $|$ $\phi \in \lbrack 0,2\pi )\}$
and admits them as eigenvectors i.e $\Phi \left\vert \phi \right\rangle
=\phi \left\vert \phi \right\rangle $. \ In terms of this operator it is
possible to express the step operators as $E_{\pm }=e^{\pm i\Phi }$ . Then a
general density matrix for the walker is written as
\begin{equation}
\rho _{W}=\int_{0}^{2\pi }\int_{0}^{2\pi }\rho (\phi ,\phi ^{^{\prime
}})\left\vert \phi \right\rangle \langle \phi ^{^{\prime }}|d\phi d\phi
^{^{\prime }}
\end{equation}%
For initial e.g $\rho _{W}=\left\vert 0\right\rangle \left\langle
0\right\vert $ we have that $\rho (\phi ,\phi ^{^{\prime }})=1.$ To start
investigating the dynamics of the walk we suppose we have initially a pure
coin density matrix $\rho _{C}=\left\vert c\right\rangle \left\langle
c\right\vert ,$ and that the first evolution step involves $k$ applications
of $V$ \ before tracing out the coin system. This choice specifies the $%
V^{k} $ quantum random walk model.\ Then the once evolved matrix $\left\vert
\phi \right\rangle \langle \phi ^{^{\prime }}|$ is $\left\vert \phi
\right\rangle \langle \phi ^{^{\prime }}|\rightarrow \varepsilon
_{V^{k}}(\left\vert \phi \right\rangle \langle \phi ^{^{\prime
}}|)=Tr_{C}(V^{k}\rho _{C}\otimes \left\vert \phi \right\rangle \langle \phi
^{^{\prime }}|V^{\dag k}),$ which is written by means $\ $\ of the Kraus
generators defined as $A_{\pm }(k,\Phi ;c)=\langle \pm |V^{k}(\Phi
)|c\rangle ,$ as follows
\begin{eqnarray}
\varepsilon _{V^{k}}(\left\vert \phi \right\rangle \langle \phi ^{^{\prime
}}|) &=&A_{+}(k,\Phi ;c)\left\vert \phi \right\rangle \langle \phi
^{^{\prime }}|A_{+}(k,\Phi ;c)^{\dag }+A_{-}(k,\Phi ;c)\left\vert \phi
\right\rangle \langle \phi ^{^{\prime }}|A_{-}(k,\Phi ;c)^{\dag }  \notag \\
&=&(A_{+}(k,\phi ;c)A_{+}(k,\phi ^{^{\prime }};c)^{\ast }+A_{-}(k,\phi
;c)A_{-}(k,\phi ^{^{\prime }};c)^{\ast })\left\vert \phi \right\rangle
\langle \phi ^{^{\prime }}|  \notag \\
&\equiv &A(k,\phi ,\phi ^{^{\prime }};c)\left\vert \phi \right\rangle
\langle \phi ^{^{\prime }}|.
\end{eqnarray}%
This means that the $n$\textit{-step evolution map} \ operates
multiplicative on $\left\vert \phi \right\rangle \langle \phi ^{^{\prime
}}|, $viz. $\varepsilon _{V^{k}}^{n}(\left\vert \phi \right\rangle \langle
\phi ^{^{\prime }}|)=A(k,\phi ,\phi ^{^{\prime }};c)^{n}\left\vert \phi
\right\rangle \langle \phi ^{^{\prime }}|,$ with $A(k,\phi ,\phi ^{^{\prime
}};c)$ to be referred to as the \textit{characteristic function } of the
walk. Then the walker density matrix evolves as
\begin{eqnarray}
\varepsilon _{V^{k}}^{n}(\rho _{W}) &=&\varepsilon _{V^{k}}^{n-1}\left(
\sum_{i=\pm }A_{i}(k,\Phi ;c)\rho _{W}A_{i}(k,\Phi ;c)^{\dagger }\right)
\notag \\
&=&\int_{0}^{2\pi }\int_{0}^{2\pi }\rho (\phi ,\phi ^{^{\prime }})A(k,\phi
,\phi ^{^{\prime }};c)^{n}\left\vert \phi \right\rangle \langle \phi
^{^{\prime }}|d\phi d\phi ^{^{\prime }}.
\end{eqnarray}%
Next is possible to evaluate the quantum moment of the walker position
operator $L,$ for the $n$ - step evolved density matrix
\begin{eqnarray}
\langle L^{s}\rangle _{n} &\equiv &Tr(L^{s}\varepsilon _{V^{k}}^{n}(\rho
_{W}))=\frac{1}{2\pi i^{s}}\int_{0}^{2\pi }d\phi \partial _{\phi }^{s}\left[
\rho (\phi ,\phi ^{^{\prime }})A^{n}(k,\phi ,\phi ^{^{\prime }};c)\right]
|_{\phi ^{^{\prime }}=\phi }  \notag \\
&=&\sum_{m\in Z}m^{s}P_{m}^{(n)}\equiv \langle m^{s}\rangle _{n}.
\end{eqnarray}%
In the last equation $P_{m}^{(n)}=\langle m|\varepsilon _{V^{k}}^{n}(\rho
_{W})|m\rangle $ \ is the classical probability for the walker to be in the
position $m$ after $n$ evolution steps ,and $\langle m^{s}\rangle _{n}$ are
the classical statistical moments of the walker position. The asymptotic
behavior of these moments for large $n$ is
\begin{equation}
\langle L^{s}\rangle _{n}=\langle m^{s}\rangle _{n}=\frac{n^{s}}{2\pi i^{s}}
\int_{0}^{2\pi } h(2\phi ;t)d\phi \left[ \rho (\phi ,\phi ^{^{\prime
}})\left( \frac{\partial }{\partial \phi }A(k,\phi ,\phi ^{^{\prime
}};c)\right) ^{s}\right] |_{\phi ^{^{\prime }}=\phi }+O(n^{s-1}).
\end{equation}%
Hence $\frac{m}{n}$ converges weakly to $h(\phi ;k,c)=-i\left[ \frac{%
\partial }{\partial \phi }A(k,\phi ,\phi ^{^{\prime }};c)\right] |_{\phi
^{^{\prime }}=\phi },$ and $\phi $ assumes the role of a random variable
with probability measure $\frac{\rho (\phi ,\phi )}{2\pi }.$ An alternative
expression for the function $h$ useful in our further investigations is
given below
\begin{equation}
h(\phi ;k,c)=Tr(\left( \sigma +V(\phi )^{\dagger }\sigma V(\phi )+...+V(\phi
)^{\dagger (k-1)}\sigma V(\phi )^{k-1}\right) \rho _{C}),
\end{equation}%
where $\sigma \equiv U_0^{\dagger }\sigma _{3}U_0.$
\begin{figure}[h]
\begin{center}
\includegraphics[height=10cm, width=11cm]{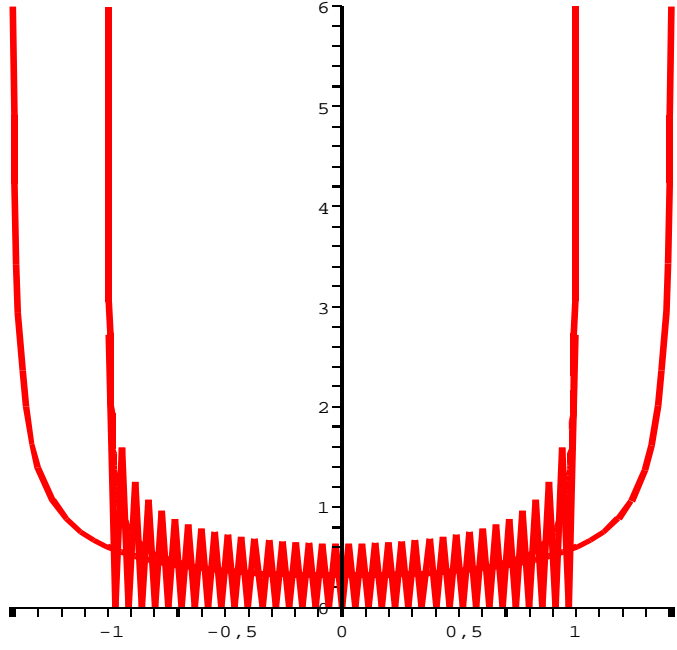}
\end{center}
\caption{ The double horn shaped limit probability distribution
function for the QRW of the $V^{2}$ model is given in the plot. Also
superimposed to it are given the occupation probabilities of the
walk as evaluated after $n=36$ evolution steps. The more spread
distribution in the graph is the distribution in \cite{konno2}
rescaled appropriately for comparison.} \label{fig1}
\end{figure}

Proceeding to evaluate the asymptotic pdf of the walk we need to resort to
the concept of the dual of completely positive trace preserving map\cite%
{kraus}. Consider the dual map $\varepsilon _{V^{k}}^{\ast }$ $%
:B(H_{W})\rightarrow B(H_{W}),$ defined on the set of bounded operators
acting on the walker Hilbert space $H_{W},$ of some given CPTP map $\
\varepsilon _{V^{k}}:D(H_{W})\rightarrow D(H_{W}),$ operating on the density
matrices $\rho _{W}\in D(H_{W})$, with operator sum realization $\varepsilon
_{V^{k}}(\rho _{W})=\sum_{i=\pm }A_{i}(k,\Phi ;c)\rho _{W}A_{i}(k,\Phi
;c)^{\dagger }.$ This dual map is defined to act on bounded operators $X\in
B(H_{W}),$ as $\varepsilon _{V^{k}}^{\ast }(X)=\sum_{i=\pm }A_{i}(k,\Phi
;c)^{\dagger }XA_{i}(k,\Phi ;c)$ . By virtue of the last definition the
expectation value (quantum moments) of the scaled powers of position
operator $(\frac{L}{n})^{s},$ evaluated after $n$ steps of the walker which
is now in the state $\rho _{W}^{(n)}=\varepsilon _{V^{k}}^{n}(\rho _{W}),$
become $\langle (\frac{L}{n})^{s})\rangle _{n}=Tr\left( \rho _{W}^{(n)}(%
\frac{L}{n})^{s}\right) ,$ or dually is determined by the equation $\langle (%
\frac{L}{n})^{s}\rangle _{n}\equiv \langle \varepsilon _{V^{k}}^{\ast n}((%
\frac{L}{n})^{s})\rangle _{0}\equiv Tr\left( \rho _{W}\text{ }\varepsilon
_{V^{k}}^{\ast n}((\frac{L}{n})^{s})\right) .$ Specializing to the $V^{2}$
model, and taking the case where initially $\rho _{W}=\left\vert
0\right\rangle \left\langle 0\right\vert ,$ and $\rho _{C}=\left\vert
c\right\rangle \left\langle c\right\vert ,$ with $\left\vert c\right\rangle
=\cos \chi \left\vert +\right\rangle +i\sin \chi \left\vert -\right\rangle ,$
we get the limit
\begin{equation}
\lim_{n\rightarrow \infty }\left\langle \varepsilon _{V^{2}}^{\ast n}\left( (%
\frac{L}{n})^{s}\right) \right\rangle _{0}=\int_{0}^{2\pi }h(\phi )^{s}\frac{%
d\phi }{2\pi }=\int_{-1}^{1}\frac{y^{s}}{\pi \sqrt{1-y^{2}}}dy.
\end{equation}%
The resulting value of the quantum moment is seen in the last equation to be
given as a statistical moment of the random variable $y^{s}$ with respect to
the limit pdf determined as follows
\begin{equation}
P(y)=\frac{1}{\pi \sqrt{1-y^{2}}},\text{ }-1\leq y\leq 1.  \label{pdf_1}
\end{equation}%
This pdf, which is the same as that obtained in \cite{kohno3},
determines asymptotically the occupation probabilities for the
scaled position variable of our QRW. In figure (\ref{fig1}) we
present its graph which has the shape of a double horn peaked at the
position $y=\pm 1.$ This is very much in difference with the
Gaussian shape of the limit pdf that occurs in a classical random
walk, but shares the double horn shape with the limit pdf of another
model of QRW\cite{konno},\cite{grimmett}, although the two
distributions differ in their exact functional form. Also in the
figure we have included the position occupation probabilities of the
QRW after a large number of steps, in order to show their tendency
towards
their asymptotic values. 


\section{ QRW Driven by Quantum Optical Cavity}

We consider now an optical cavity where an interaction between a single
electromagnetic (EM) mode and a single two-level atomic system takes place
on resonance. Four different types of this interaction will be considered
giving rise to four respective solvable models. These are the
Jaynes-Cummings model (JCM) $H_{JCM},$ the intensity dependent JCM $%
H_{ID-JCM},$ the two photon JCM $H_{2ph-JCM},$ and the $m-$photon JCM $%
H_{mph-JCM},$ with the corresponding Hamiltonians given below:

\begin{eqnarray}
H_{JCM} &=&\frac{\omega }{2}\sigma _{3}+\omega a^{\dagger }a+\lambda (\sigma
^{+}a+\sigma ^{-}a^{\dagger }), \\
H_{ID-JCM} &=&\frac{\omega }{2}\sigma _{3}+\omega a^{\dagger }a+\lambda
(\sigma ^{+}a\sqrt{N}+\sigma ^{-}a^{\dagger }\sqrt{N+1}), \\
H_{2ph-JCM} &=&\frac{\omega }{2}\sigma _{3}+\omega a^{\dagger }a+\lambda
(\sigma ^{+}a^{2}+\sigma ^{-}a^{\dagger 2}),\text{ and} \\
H_{mph-JCM} &=&\frac{\omega }{2}\sigma _{3}+\omega a^{\dagger }a+\lambda
(\sigma ^{+}a^{m}+\sigma ^{-}a^{\dagger m}).
\end{eqnarray}

In the above expressions $a^{\dagger },$ $a$ \ are the creation,
annihilation operators of quanta of the EM field satisfying the canonical
commutation relation $[a,a^{\dagger }]=\mathbf{1}$, while the $\sigma
^{+},\sigma ^{-}$ realize the step operators in the atomic "quantum coin"
system state space, which together with $\sigma _{3}$ satisfy the Pauli
matrix algebra $[\sigma _{+},\sigma _{-}]=2\sigma _{3},$ $[\sigma
_{3},\sigma _{\pm }]=\pm \sigma _{\pm }$ . Also $\lambda $ is the field-atom
coupling constant and $\omega $ stands for the atomic energy difference
which equals the field frequency in the case of resonance, as in our case.

The dynamic symmetry of the above Hamiltonians can be seen by writing them
as $\ H_{k}=C+V_{k},$ $k=$JCM,ID-ICM, $2$ph-JCM, mph-JCM, where $C= \frac{
\omega }{2}\sigma _{3}+\omega N,$ with $N=a^{\dagger }a$ the number
operator, is the common free part of the Hamiltonians and represents the
total "number of excitations" and the $V_{k}$ obviously identified with the
rest part of the Hamiltonian stands for the interaction between atom and the
field. We can verify the following constants of motion: $[H_{k},C]=0,$ $%
[H_{k},V_{k}]=0,$ $[C,V_{k}]=0.$

In the interaction picture the unitary evolution operator reads for each of
these models $U_{k}(t)=\exp (-itV_{k}).$ Explicitly we obtain
\begin{eqnarray}
U_{JCM}(t) &=&\left(
\begin{array}{cc}
\cos (\lambda t\sqrt{aa^{\dagger }}) & -ia\frac{\sin (\lambda t\sqrt{%
a^{\dagger }a})}{\sqrt{a^{\dagger }a}} \\
-ia^{\dagger }\frac{\sin (\lambda t\sqrt{aa^{\dagger }})}{\sqrt{aa^{\dagger }%
}} & \cos (\lambda t\sqrt{a^{\dagger }a}%
\end{array}
\right) ,\text{ \ } \\
U(t)_{ID-JCM} &=&\left(
\begin{array}{cc}
\cos (\lambda t\text{ }aa^{\dagger }) & -ia\sqrt{N}\frac{\sin (\lambda t%
\text{ }a^{\dagger }a)}{a^{\dagger }a} \\
-i\sqrt{N}a^{\dagger }\frac{\sin (\lambda t\text{ }aa^{\dagger })}{%
aa^{\dagger }} & \cos (\lambda ta^{\dagger }a)%
\end{array}
\right)
\end{eqnarray}
and for the $m-$photon case
\begin{equation}
U_{mph-JCM}(t)=\left(
\begin{array}{cc}
\cos (\lambda t\sqrt{a^{m}a^{\dagger m}}) & -ia^{m}\frac{\sin (\lambda t%
\sqrt{a^{\dagger m}a^{m}})}{\sqrt{a^{\dagger m}a^{m}}} \\
-ia^{\dagger m}\frac{\sin (\lambda t\sqrt{a^{m}a^{\dagger m}})}{\sqrt{%
a^{m}a^{\dagger m}}} & \cos (\lambda t\sqrt{a^{\dagger m}a^{m}})%
\end{array}
\right) ,
\end{equation}
which is specialized to the case of two-photonic transition when
$m=2.$ Assuming the initial field state is the pure state $\rho
_{f}=|f\rangle \langle f|,$ and the atomic coin state the general
density matrix $\rho _{C}, $ the state of the atomic system after
its crossing through the first quantum optical cavity is now
described by the reduced density matrix given below which is obtain
by tracing out the field degree of freedom namely,
\begin{equation}
\varepsilon _{U}(\rho _{C})=Tr_{f}\left( U(t)(\rho _{C}\bigotimes
\rho _{f})U(t)^{\dagger }\right) .
\end{equation}%
Introducing the operators $P_{ij}=|i\rangle \langle j|$ in the quantum coin
space$,$ then the map
\begin{equation}
\varepsilon _{U}(\rho _{C})=\sum_{ij,kl=0,1}P_{ij}\rho _{C}P_{kl}\langle
f|U_{kl}^{\dagger }U_{ij}|f\rangle ,
\end{equation}
is a positive and trace preserving transformation of the atomic density
matrix i.e if $\rho _{C}>0,$ then $\varepsilon _{U}(\rho _{C})>0,$ and also $%
Tr(\varepsilon _{U}(\rho _{C}))=Tr\rho _{C},$ as is easily shown. For the
particular case of the JCM with the field been initially in the vacuum state
i.e $|f\rangle =|0\rangle ,$ we obtain the map
\begin{equation}
\varepsilon _{U}(\rho _{C})=S_{0}\rho _{C}S_{0}^{\dagger }+S_{1}\rho
_{C}S_{1}^{\dagger }
\end{equation}%
where the so called Kraus generators of that map are
\begin{equation}
S_{0}(t)=\left(
\begin{array}{cc}
\cos (\lambda t) & 0 \\
0 & 1%
\end{array}%
\right) ,\text{ }S_{1}(t)=\left(
\begin{array}{cc}
0 & 0 \\
\sin (\lambda t) & 0%
\end{array}%
\right) ,
\end{equation}%
and satisfy the property $S_{0}^{\dagger }S_{0}+S_{1}^{\dagger }S_{1}=%
\mathbf{1.}$

The generalization of this result to the case of all four models when the
field is in some sharp number state $|f\rangle =|r\rangle $ $r=0,1,2,...,$
leads to the reduced coin density matrix
\begin{equation}
\varepsilon _{U}(\rho _{C})=A_{1}\rho _{C}A_{1}^{\dagger }+A_{2}\rho
_{C}A_{2}^{\dagger }+A_{3}\rho _{C}A_{3}^{\dagger }
\end{equation}%
with Kraus generators
\begin{eqnarray}
A_{1} &=&\left(
\begin{array}{cc}
\cos (\lambda t\eta ) & 0 \\
0 & \cos (\lambda t\theta )%
\end{array}%
\right) ,\text{ }A_{2}=\left(
\begin{array}{cc}
0 & 0 \\
\sin (\lambda t\eta ) & 0%
\end{array}%
\right) ,\text{ }  \notag \\
A_{3} &=&\left(
\begin{array}{cc}
0 & \sin (\lambda t\theta ) \\
0 & 0%
\end{array}%
\right) ,
\end{eqnarray}
where $\eta =$
$\sqrt{r+1},r+1,\sqrt{(r+1)(r+2)},\sqrt{\frac{(r+m)!}{r!}}$\ and
$\theta =\sqrt{r},r,r(r-1),$ $\sqrt{\frac{r!}{(r-m)!}},$ are the
angles
for the respective models viz. the JCM, the ID-JCM, $2$-photon, and the $m$%
-photon JCM. The trace preservation of this map requires that $%
A_{1}^{\dagger }A_{1}+A_{2}^{\dagger }A_{2}+A_{3}^{\dagger }A_{3}=\mathbf{1}%
. $

\section{ Cavity Driven QRW Statistics}

The effect of the first optical cavity will be to transform $\rho
_{C}=\left\vert c\right\rangle \left\langle c\right\vert $ to $\varepsilon
_{U}(\rho _{C}),$ this yields for the particular choice of $|c\rangle =\cos
\chi |+\rangle +i\sin \chi |-\rangle ,$ the mixed coin state
\begin{eqnarray}
\varepsilon _{U}(\rho _{C}) &=&\frac{1}{2}\mathbf{1}+\frac{1}{2}\sin (2\chi
)\cos \left( \ \lambda \eta t\right) \cos \left( \ \lambda \theta t\right)
\sigma _{2}  \notag \\
&&+\frac{1}{2}\left[ \cos \left( 2\lambda \eta t\right) \cos ^{2}\chi -\cos
\left( 2\lambda \theta t\right) \sin ^{2}\chi \right] \sigma _{3}.
\end{eqnarray}%
In this \ case the characteristic function is time dependent and reads
\begin{eqnarray}
h(\phi ;t) &=&\left[ -\cos \left( 2\lambda t\eta \right) \cos ^{2}\chi +\cos
\left( 2\lambda t\theta \right) \sin ^{2}\chi \right] \cos (2\phi )  \notag
\\
&&+\left[ \sin (2\chi )\cos \left( \ \lambda t\eta \right) \cos \left( \
\lambda t\theta \right) \right] \sin (2\phi ).  \label{h_2}
\end{eqnarray}

We now proceed to evaluate the limit probability distribution function, and
to this end we rewrite the last equation as $h(\phi ;t)=C(t)\cos [2\phi
-\Lambda (t)]$, where we have introduce the functions $A(t)=-\cos \left(
2\lambda t\eta \right) \cos ^{2}\chi +\cos \left( 2\lambda t\theta \right)
\sin ^{2}\chi ,$ $B(t)=\sin (2\chi )\cos \left( \ \lambda t\eta \right) \cos
\left( \ \lambda t\theta \right) $ from which we define the two functions $%
C(t)=\sqrt{A(t)^{2}+B(t)^{2}}$, \ and $\tan \Lambda (t)=\frac{B(t)}{A(t)}$.
\ If we now set $y=h(\phi ;t)$ then for all four inverses $h_{i}^{-1},$ $%
i=1,2,3,4,$ we get that $h^{\prime
}(h_{i}^{-1}(y))=-2\sqrt{C(t)^{2}-y^{2}}$. This results into the limit pdf which is
$P(y;t)=\frac{1}{2\pi }%
\sum\limits_{i}\frac{1}{|h^{\prime }(h_{i}^{-1}(y))|}$, or finally
\begin{equation}
P(y;t)=\frac{1}{\pi \sqrt{C(t)^{2}-y^{2}}},\text{ }-1\leq y\leq 1.
\label{pdf_2}
\end{equation}

The influence of the first cavity is now expressed by the double
dependence of the distribution, first on the time spent in the
cavity i.e the coin-field interaction time, and second on the
initial coin state by means of the dependence of $C(t),$ via $A(t)$
and $B(t),$ on the angle $\chi $ of the coin vector. The
relationship
\begin{equation}
\label{distr}
\int_{-1}^{1}\frac{dy}{\sqrt{1-y^{2}}}=\int_{-|C(t)|}^{|C(t)|}\frac{dy}{\sqrt{%
C(t)^{2}-y^{2}}}
\end{equation}%
between the limit distribution of the walk without the driving cavity (eq.(%
\ref{pdf_1})), and the same one with the driving cavity in presence (eq.(\ref%
{pdf_2})), shows that when $C(t)$ is not zero the asymptotics of QRW are
robust to the changes caused by the modified coin system, up to a scaling.
The scaling explicitly refers to the changes $y(t)\rightarrow y(t)/|C(t)|,$ $%
(-1,1)\rightarrow (-|C(t)|\rightarrow |C(t))|,$ in the random
variable and its interval of values respectively.

\begin{figure}[h]
\begin{center}
\includegraphics[height=10cm, width=11cm]{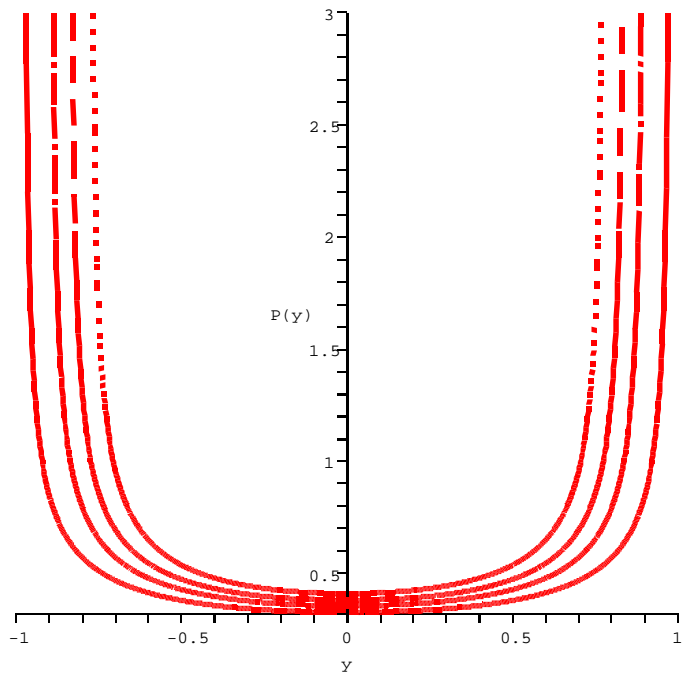}
\end{center}
\caption{ The limit pdf of fig.(\protect\ref{fig1}), as it has been affected
by the presence of the optical cavity. The parameters $t$ and $\protect\chi %
, $ are chosen so the function $C(t)^{2},$ modifying the limit pdf
(see text), takes the values $C(t)^{2}=\{0.95,0.8,0.7,0.6\}.$ The
respective plots are with solid line, dotted line, dashed line, and
dashed-dotted line. Apart from scaling, the limit pdf reappears in
the asymptotics of the walk. This scaling would lend itself to
experimental detection of the robustness of the asymptotic pdf of
the quantum walk.} \label{fig2}
\end{figure}

Next, in figure (\ref{fig2}), we give the graph of
eq.(\ref{pdf_2}) at various non zero values of $C(t).$


The statistical moment derived by the new distribution are now in general
time dependent, e.g the first two of them are found to be
\begin{eqnarray}
\mu &=&\lim_{n\rightarrow \infty }\langle \frac{L}{n}\rangle
_{n}=\int_{0}^{2\pi} h(2\phi ;t)\frac{d\phi }{2\pi }=0, \\
\sigma (t)^{2} &=&\lim_{n\rightarrow \infty }\langle \left( \frac{L}{n}
\right) ^{2}\rangle _{n}\text{ }=\int_{0}^{2\pi} h(2\phi ;t)h(\phi ;t)^{2}%
\frac{ d\phi }{2\pi }=\frac{C(t)^{2}}{2},
\end{eqnarray}
namely the first one is zero so that the walk remains unbiased while the
standard deviation depends on the time spent by the coin in the first cavity.

To further probe the behavior of the QRW and especially the
influences upon its asymptotics of the driving cavity, we first note
that for $t=0,$ that is in the absence of the first cavity, the
function $h$ depends on the initial state of the quantum coin system
by means of its angle $\chi $ i.e $h(\phi ;0)=-\cos (2\phi +2\chi
).$ However this dependence does not show up in the asymptotic
regime since its probability distribution given in eq.(\ref{pdf_1}
), appears to have a universal character, and is independent from
the initial coin state provided it is of the form considered here.

On the other hand as eqs.(\ref{h_2},\ref{pdf_2}) indicate in the case that
there is a first cavity present the dependence of function $h$ on the
initial coin state survives in the asymptotic regime for the $JCM$ and its
variations. Indeed by inspection of the functions $A(t),$ $B(t)$ and $C(t)$
as given above we see that they do depend on the $\chi $ angle, and this
dependence harbors the possibility of controlling the asymptotic statistics
of the walk. By choosing the initial coin state to have angles $\chi =\{0,%
\frac{\pi }{2},\pi ,\frac{3\pi }{2}\},$ namely to be $|c\rangle =\{|+\rangle
,i|-\rangle ,-|+\rangle ,-i|-\rangle \}$ respectively, we enforce $B(t)=0,$ $%
\forall t.$ If we further choose the interaction time in the first cavity so
that $A(t)=0,$ which implies correspondingly the times $t=\{\frac{(2k+1)\pi
}{4\lambda \eta },$ $\frac{(2k+1)\pi }{4\lambda \theta },\frac{(2k+1)\pi }{%
4\lambda \eta },$ $\frac{(2k+1)\pi }{4\lambda \theta }\},$ $k\in Z$, then we
get also $C(t)=0$ for those values of $t.$ Then the specific relations among
the four coin states and interaction times as given above result into only
two different pairs of coin density matrices and interaction times namely, $%
\rho _{C}=|+\rangle \langle +|,$ $t=\frac{(2k+1)\pi }{4\lambda \eta },$ and $%
\rho _{C}=|-\rangle \langle -|,$ $t=\frac{(2k+1)\pi }{4\lambda \theta },$
for which $C(t)=0,$ a fact that holds true for all JC models of our study.
If each of these two conditions occur we say that a \textit{resonance
condition }takes place in the first cavity between the field and the two
level atom. \ So we see that the resonance condition implies that the
standard deviation becomes zero, hence the limit pdf collapses, and more
importantly that $<L^{2}>_{n}$ $\sim n,$ so we loose the quadratic diffusion
time speed up, characterizing the quantum random walk. In such a case the
asymptotic behavior of the standard deviation agrees with that of a
classical random walk. More precisely what happens is that in all the above
cases, the exiting coin from the first cavity is in the maximally
classically mixed state $\rho _{C}=\frac{1}{2}|+\rangle \langle +|+\frac{1}{2%
}|-\rangle \langle -|$ $=\frac{1}{2}\mathbf{1}.$ If this coin system is used
to feed in the second cavity where the $V^{2}$ QRW takes place, then the
final one-step density matrix for the walker system becomes
\begin{equation}
\rho _{W}\rightarrow \varepsilon _{V^{2}}(\rho _{W})=\frac{1}{2}\rho _{W}+%
\frac{1}{4}E_{+}^{2}\rho _{W}E_{+}^{\dagger 2}+\frac{1}{4}E_{-}^{2}\rho
_{W}E_{-}^{\dagger 2}.
\end{equation}

A comment is on order. If $\rho _{W}$ is initially a diagonal matrix then so
is finally, because of the last equation. Hence we really have a classical
one-step transition that leads to Gaussian statistics for large $n,$ once we
normalize $L$ to $\frac{L}{\sqrt{n}}$ . This implies that on resonance the
walk becomes fully classical. This analysis makes obvious the fact that a
judicious choice of the initial coin state permits us to tune the
interaction time in the first cavity where the JCM or some of its variations
is implemented, so that we have an absolute control not only over the
asymptotic behavior of QRW, but on the very quantum nature of its
performance as well. This conclusion makes the quantum optical experimental
investigation of this idea worthwhile.


\section{Conclusions}

The QRW of the model $V^{2}$ exhibits a quadratically enhanced
diffusion rate compared to the rate of the classical walk, and an
interesting and counter intuitive \ limit probability distribution.
This distribution has been investigated in the present work with
respect to its behavior under variations of the quantum coin state
that are tailored in a cavity preceding the black box structure
where the QRW itself is implemented. As the quantum coin is taken to
be a two-level atom, its state alterations have been induced by
letting it interact with a quantum mode in a JCM type of
interaction on resonance. To gain generality in addition to the original JCM$%
,$ three of its versions namely the intensity dependent, the two-photon and
the $m-$ photon JCM have been used. All four models give rise to similar
alteration of the asymptotics of the walk, a fact that shows the generality
of the obtained results concerning asymptotics.

The two main modifications found of the long time limit
probabilities of the QRW, are parametrized by the state of the coin
going \textit{into} the cavity, and the time spent by the coin
\textit{in} the cavity. In the first modification, the tuning of
these two parameters to the \textit{resonance condition }leads to
fully classical results, namely to Gaussian limit distribution. In
the second modification, a complementary situation prevails in which
the double horn shape limit pdf re-emerges up to a scaling. These
phenomena are independent of the particular version of the resonant
JCM used to realize the driving cavity. Their experimental
verification seems to be feasible by the present cavity QED
experimental settings. Additional studies about e.g. the role of
interaction time variations, the decoherence due to spontaneous
emission of quantum coins, and the statistics of the arrival times
of \ quantum coins in the cavity, are necessary, and will be taken
up elsewhere.

\end{document}